\documentclass{PoS}

\title{Sensitivity to Atypical Tau Initiated Air Showers For a High-Altitude Optical Cherenkov Detector}

\ShortTitle{ANITA events in Optical Cherenkov}

\author{\speaker{A. Cummings} $^{1,2 \dag}$, R. Aloisio$^{1,2}$, M. Bertaina$^{3,4}$, F. Bisconti$^{3,4}$, F. Fenu$^{3,4}$, F. Salamida$^{2,5}$\\
\llap{$^1$}Gran Sasso Science Institute, L'Aquila, Italy\\
\llap{$^2$}INFN, Laboratori Nazionali Gran Sasso, Assergi (L'Aquila), Italy\\
\llap{$^3$}Universita di Torino, Torino, Italy\\
\llap{$^4$}INFN Torino, Torino, Italy\\
\llap{$^5$}Universita dell'Aquila, Dipartimento di Scienze Fisiche e Chimiche, L'Aquila, Italy\\
\llap{$^\dag$}E-mail: \email{austinlee.cummings@gssi.it}}

\abstract{The ANITA collaboration has recently announced the supposed observation of two upward going cosmic ray showers at earth emergence angles $27^{\circ}$ and $35^{\circ}$ with reconstructed energy $\sim$0.6~EeV. Upward air showers (UAS) are expected from tau leptons resulting from the interaction of astrophysical neutrinos inside the Earth. However, at emergence angles larger than $20^{\circ}$, the probability of tau emergence from a  neutrino is less than $10^{-7}$, which makes a standard model explanation for these signals difficult.\\

If confirmed by other experiments, these energetic events would strengthen the argument for physics beyond the standard model. Both the proposed EUSO-SPB2 and the POEMMA instruments will be equipped with optical Cherenkov detectors in order to measure the Cherenkov emission from UAS, which, if aimed low enough below the horizon, could, in principle, capture these events as well. An observation in the Cherenkov channel would help to rule out anthropogenic and other explanations for these events. We present here the sensitivity to the ANITA anamolous events for a balloon based and a satellite based Cherenkov detector, as could be realized in the upcoming EUSO-SPB2 mission and the proposed POEMMA mission, respectively.}

\FullConference{36th International Cosmic Ray Conference -ICRC2019-\\
		July 24th - August 1st, 2019\\
		Madison, WI, U.S.A.}

\begin{document}

\section{Introduction}
Tau neutrinos are produced in oscillations of cosmic neutrinos as they travel from their sources to Earth. These neutrinos will produce a flux of tau leptons after propagation through the Earth, where they undergo charged and neutral current interactions, as well as energy losses, and possible decay/regeneration. The exiting tau lepton can decay in the atmosphere, producing an upward moving extensive air shower, referred to as a UAS.

\begin{wrapfigure}{L}{0.6\textwidth}
\includegraphics[width=\linewidth]{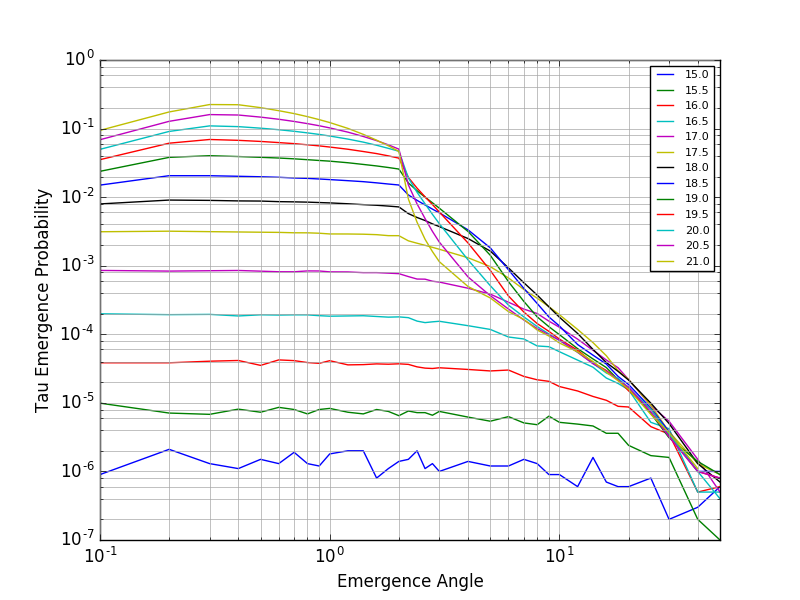}
\caption{Tau emergence probability as a function of Earth emergence angle for different primary neutrino energies}
\label{Tau_Probs}
\end{wrapfigure}

ANITA is a balloon based radio telescope, which has made several flights over the Antarctic ice sheet. ANITA measures radio pulses from 1) conventional downward going showers reflected off the ice 2) upward going cosmic ray showers from above the horizon 3) Askaryan radiation from neutrino interactions in ice, but most importantly for this study: 4) UAS initiated by tau leptons resulting from neutrino propagation through the Earth. ANITA has recently measured 2 signals that could be explained as UAS, with energies close to $\sim$ 0.6~EeV and Earth emergence angles $27^{\circ}$ and $35^{\circ}$ \cite{ANITA} \cite{ANITA2}. However, these observations seem unlikely when taking into account standard model tau emergence probabilities as shown in figure \ref{Tau_Probs}.

The emergence probability of a tau lepton with parent neutrino energies greater than $10^{17}$~eV at Earth emergence angles $27^{\circ}$ and $35^{\circ}$ is $8 \times 10^{-6}$ and $2 \times 10^{-6}$, respectively. Moreover, there are two processes to consider which lower the individual probability of these events considerably: the energy distributions of tau lepton decay products which may initiate an air shower and the energy distribution of emerging taus due to energy losses inside the Earth.

\begin{wrapfigure}{R}{0.6\textwidth}
\includegraphics[width=\linewidth]{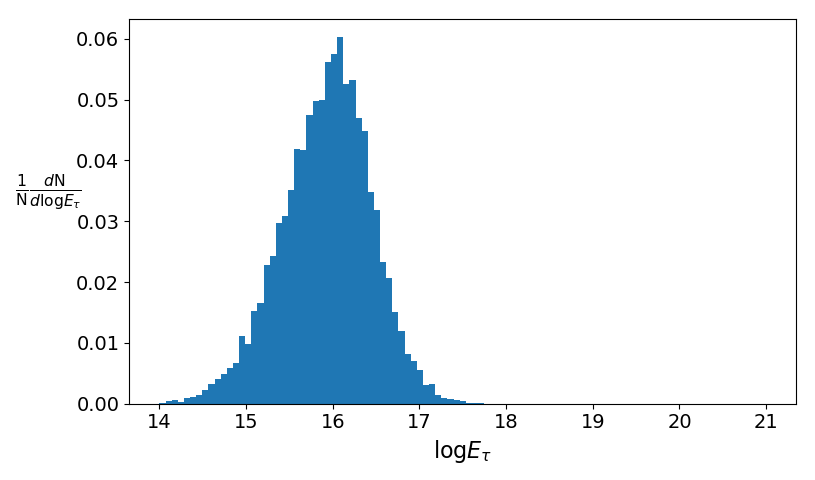}
\caption{Energy distribution of exiting tau lepton from a $10^{21}$~eV tau neutrino at $27^{\circ}$ Earth emergence angle}
\label{Energy_Dists}
\end{wrapfigure}

The average fractional energy of a tau lepton which goes into hadronic or electronic decay products which may conventionally shower in the atmosphere is roughly $50\%$, which, for the ANITA-like events, indicates a parent tau lepton energy of $\sim$1~EeV. Figure \ref{Energy_Dists} represents the energy distribution of an emerging tau lepton from a $10^{21}$~eV parent tau neutrino at $27^{\circ}$ Earth emergence angle. Figures \ref{Tau_Probs} and \ref{Energy_Dists} were generated using a custom tau neutrino monte carlo propagation code. From the distribution, we can fit an approximate Gaussian in log energy space, with mean $\mu = 15.86$ and RMS $\sigma = 0.50$. The probability that an event from this distribution has energy $10^{18}$~eV is roughly $2 \times 10^{-5}$. We note that using the distribution for a $10^{21}$~eV neutrino is the most optimistic case for obtaining an emerging 1~EeV tau. We therefore calculate that the upper bound probability for a tau neutrino to develop showers with the energy and geometry detected by ANITA under standard model assumptions would be between $1.6 \times 10^{-10}$ and $4 \times 10^{-11}$.

Assuming for granted that the interpretation of the ANITA events as upgoing showers initiated by tau leptons, it is reasonably certain that it is not a result of standard model processes, which makes these events interesting. However, ANITA observation alone cannot confirm nor reject beyond standard model explanations for these signals. Future experiments such as EUSO-SPB2 and POEMMA seek to measure the upward air showers induced by tau leptons by observing the Cherenkov light produced from the charged particles in the shower from high altitude and space, respectively \cite{EUSOSPB2} \cite{POEMMA}. A valid detection or non-detection in the Cherenkov channel would help constrain the explanations for these events.

In the framework of standard model physics, one possible interpretation of the ANITA events is that standard radio emission from cosmic rays are reflected off unique subsurface features of the Antarctic ice, thereby avoiding the polarity reversal which conventional, downward going, reflected cosmic ray air showers exhibit \cite{explanation1}. Another standard model explanation is that coherent transition radiation allows for an altered electromagnetic field at the air-ice transitional boundary, thereby allowing showers with small zenith angles to produce non-inverted signals \cite{explanation2}. Both of these SM explanations are based on the idea of a downward going shower appearing as an upward shower through various means. Observations in the Cherenkov emission band would be useful on the distinction between upward and downward going showers. Compared to those in an upward going shower, Cherenkov photons from a downward going shower would experience additional losses from 1) reflection on ice 2) tranversing the path length in the atmosphere twice 3) having increased space for lateral spreading. Thus, there should be a significant difference in the Cherenkov intensity between upward and downward going showers which can be observed. In this work, we determine what the sensitivity of EUSO-SPB2 and POEMMA instruments would be to upward going ANITA like events in the Cherenkov channel and explore whether they could be valid tools for confirm or reject the ANITA claim.

\section{Geometry}
The tau lepton decays on the scale $c \tau E_{\tau}/m_{\tau}$ = $4.9 (E_{\tau}/10^{17}~\mathrm{eV})$~km. For the two ANITA events, this translates to an average decay length of about 50~km. In a spherical geometry, path length is calculated via:

\begin{equation}
L(z,\theta) = \sqrt{R_{E}^{2} \mathrm{cos}^{2}\left(\frac{\pi}{2}-\theta \right)+z^{2}+2zR_{E}}-R_{E}\mathrm{cos} \left(\frac{\pi}{2}-\theta \right)
\end{equation}

Where $R_{E}$ is the Earth's radius and $\theta$ is the emergence angle of the tau lepton. The first EUSO-SPB experiment flew at 33~km, which corresponds to a path length of 72~km and 57~km for the two geometries. Assuming that EUSO-SPB2 will fly at a comparable altitude, the fraction of events which decay within this range is $1-e^{-72(57)\mathrm{km}/L}$, or $76\%$($68\%$). For POEMMA, which will orbit at an altitude of 525~km, this is very close to $100\%$. Thus, the ANITA events will decay within the discovery range of EUSO-SPB2 and POEMMA. For EUSO-SPB2, it is likely that the instrument will be inside the shower development \cite{EUSOSPB2} \cite{POEMMA}.

We must also consider that both POEMMA and EUSO-SPB2 are not designed exclusively for Cherenkov light collection. Both experiments include two different cameras. The largest camera looks directly downwards and is designed to image the fluorescence emission from ultra-high energy cosmic ray extensive air showers. The Cherenkov camera, however, views only a small portion of the Earth's limb, where neutrinos have a high probability of producing showering tau leptons. The Cherenkov camera of EUSO-SPB2 has an optical field of view $45^{\circ} \times 3.5^{\circ}$, and is designed for rotation, such that it can view, at maximum, $10^{\circ}$ below the horizon, which corresponds to a maximum observable Earth emergence angle of $15^{\circ}$. POEMMA is designed such that it will view up to a fixed $7^{\circ}$ below the horizon, which corresponds to $20^{\circ}$ maximum observable Earth emergence angle. Neither angular range covers the reconstructed angles of the ANITA events. In order for EUSO-SPB2 and POEMMA to have sensitivity in this region of Earth emergence angles, design changes must be made to increase the effective field of view. These design changes are within reach of the EUSO-SPB2 instrument, as the Cherenkov camera is already designed for angular rotation. Conversely, for POEMMA, the field of view would need to be increased by a factor of 2-3. A detailed discussion of the optimal configuration of a balloon or satellite based Cherenkov detector will be discussed elsewhere \cite{me}.

%In order for EUSO-SPB2 to have sensitivity in this region, it would need to view between $22^{\circ}$ and $30^{\circ}$ below the horizon. For POEMMA, these values are $12^{\circ}$ and $17^{\circ}$. There is not a strong justification to increase the angular range for POEMMA by a factor of 2. Doing so would measurably increase cost, and the conventional tau lepton flux is extremely low for larger emergence angles, so there is little other reason to explore this range. However, the EUSO-SPB2 Cherenkov camera is being designed to rotate, such that it can measure direct emission from cosmic rays above the horizon. Perhaps it makes sense to increase the range of rotation, with the goal of seeking these events.

\section{Simulations}
We developed a customized Cherenkov propagation code, which calculates the spectrum, spatial distribution, and intensity of Cherenkov photons at a fixed altitude, taking into account the inherent spherical geometry of the curved Earth atmosphere, given as input the charged particle longitudinal distribution. The electron angular distribution used is taken from \cite{electrons}, and the atmospheric transmission is described in \cite{Atmosphere}. The charged particle profile can be given from either a particle propagation code, or from an arbitrary parameterization \cite{me}.

 To generate the charged particle profiles needed as input for the Cherenkov simulation, we use a modified version of CORSIKA-75600 \cite{corsika}, which allows us initiation of showers at true ground level. The geometry we use for our simulations is upgoing, and nearly perfectly horizontal (to allow for a sufficient upper bound on the total slant depth to use in longitudinal profiles). We have also modified the decay times of tau leptons in CORSIKA such that they decay immediately. In this manner, we monitor the shower induced by the tau decay products without the possibility of decay outside the atmosphere. This is also to say that we ignore the ionization energy losses of the tau lepton through air, as it is a very small effect. We later adjust the profiles by an appropriate tau decay distance in the atmosphere, and account for the proper Earth emergence angle of the tau. The atmospheric model we use is that of the US Standard Atmosphere \cite{Atmosphere}.
 
 In this section, we detail the properties of the arriving Cherenkov light from an event with ANITA like properties as would be observed by POEMMA and EUSO-SPB2, and discuss whether they are within the potential range for discovery. For the following calculations, we use the 1st ANITA event (with $27^{\circ}$ Earth emergence angle), as it would be the geometry easier to observe under the described configuration. For the spectra and spatial distribution calculations, we use a proton shower with varying first interaction point to demonstrate the average behavior of the tau.

\subsection{Cherenkov Spectra}
%The Cherenkov spectra for a 100~PeV proton with ANITA-like geometry is plotted in figure \ref{Cherenkov_Spectra} with decay distances ranging from 0~km to 45~km. As the shower development begins deeper in the atmosphere, the peak wavelength of the distribution decreases, due to the decreased atmospheric attenuation, which is more prominent for smaller wavelengths. In a  similar manner, the total intensity increases with increasing first interaction point because of decreased atmospheric attenuation and geometric spreading, up until an altitude where the charged particles in the shower no longer reach the Cherenkov threshold, and cannot efficiently generate photons. Also note that the increase in signal at low wavelength seen by POEMMA but not by EUSO-SPB2 is due to shower development which occurs outside the view of SPB (that is, above 33~km altitude), which has very little atmospheric attenuation at high altitudes. The Cherenkov camera of both instruments will be sensitive in the wavelength range 300~nm-550~nm. This range covers over $60\%$ of the total photons, and includes the peak intensity for any decay distance of the tau lepton.

\begin{wrapfigure}{L}{0.5\textwidth}
\includegraphics[width=\linewidth]{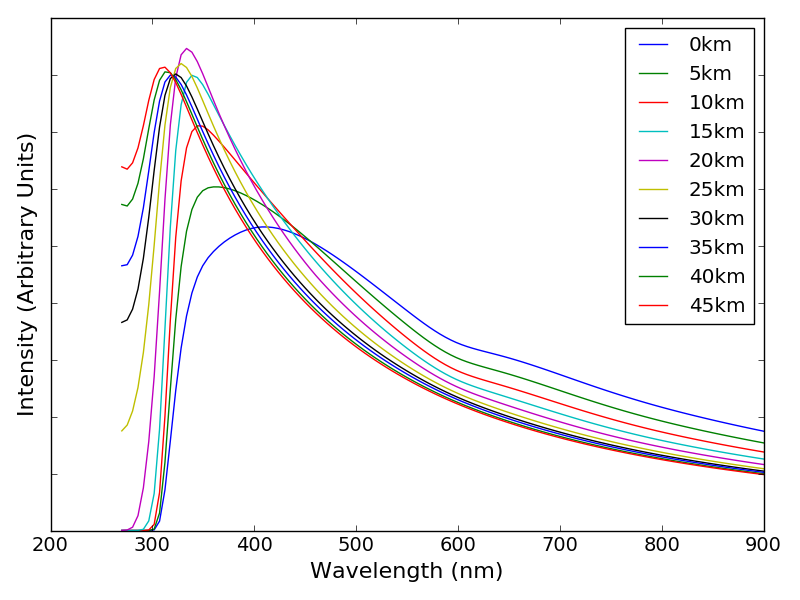}
 \caption{Normalized Cherenkov spectra for a 100~PeV proton shower with $27^{\circ}$ Earth emergence angle (ANITA event 1) initiated at different decay distances with wavelength range 270~nm to 900~nm, using the atmosphere given by Elterman \cite{Atmosphere}}
\label{Cherenkov_Spectra}
\end{wrapfigure}

The normalized Cherenkov spectra for a 100~PeV proton with ANITA-like geometry is plotted in figure \ref{Cherenkov_Spectra} with decay distances ranging from 0~km to 45~km. As shower development begins deeper in the atmosphere, the peak wavelength of the distribution decreases, due to the decreased atmospheric attenuation, which is more prominent for smaller wavelengths. The Cherenkov camera of both instruments will be sensitive in the wavelength range 300~nm-550~nm. This range covers over $60\%$ of the total photons, and includes the peak intensity for any decay distance of the tau lepton.

%\begin{figure}[t!]
%\begin{tabular}{cc}
%  \includegraphics[width=.5 \textwidth]{Spectrum_SPB.png} %& 
%  \includegraphics[width=.5 %\textwidth]{Spectrum_POEMMA.png}
%
%\end{tabular}
% \caption{Cherenkov spectra for a 100~PeV proton shower with $27^{\circ}$ Earth emergence angle (ANITA event 1) initiated at different decay distances for EUSO-SPB2 (left) and POEMMA (right) from 270~nm to 900~nm, using the atmosphere given by Elterman \cite{Atmosphere}}
%\label{Cherenkov_Spectra}
%\end{figure}

\subsection{Spatial Distribution}
\begin{wrapfigure}{R}{0.5\textwidth}
\includegraphics[width=\linewidth]{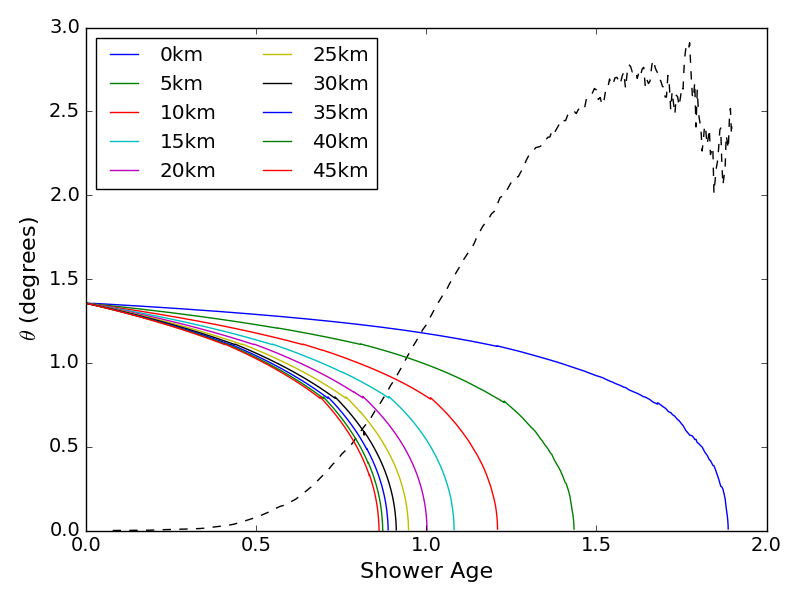}
\caption{Electron scale angle (dashed) versus Cherenkov angle as a function of shower age for decay distances between 0~km and 50~km for a 100~PeV proton shower with $27^{\circ}$ Earth emergence angle.}
\label{Scale_Angle}
\end{wrapfigure}

The spatial distribution of Cherenkov photons is dependent on 3 factors: the Cherenkov angle which varies as a function of altitude, the electron angular distribution, which varies as a function of shower age ($s = 3/(1+2(X_{\mathrm{max}}/X))$), and the distance from the emission to the detector plane. For a given energy, the electron angular distribution is described well by an exponential $e^{-\theta/\theta_{\mathrm{s}}}$, described in \cite{electrons}.  In figure \ref{Scale_Angle}, we plot the average scale angle for an upwards 100~PeV proton shower in the ANITA geometry (although we note that this is independent of primary energy \cite{Lipari}) against the corresponding Cherenkov angle as a function of shower age. From figure \ref{Scale_Angle}, we can estimate the dominant behavior of the Cherenkov photon distribution for different decay distances.

\begin{figure}[t!]
\begin{tabular}{cc}
  \includegraphics[width=.5 \textwidth]{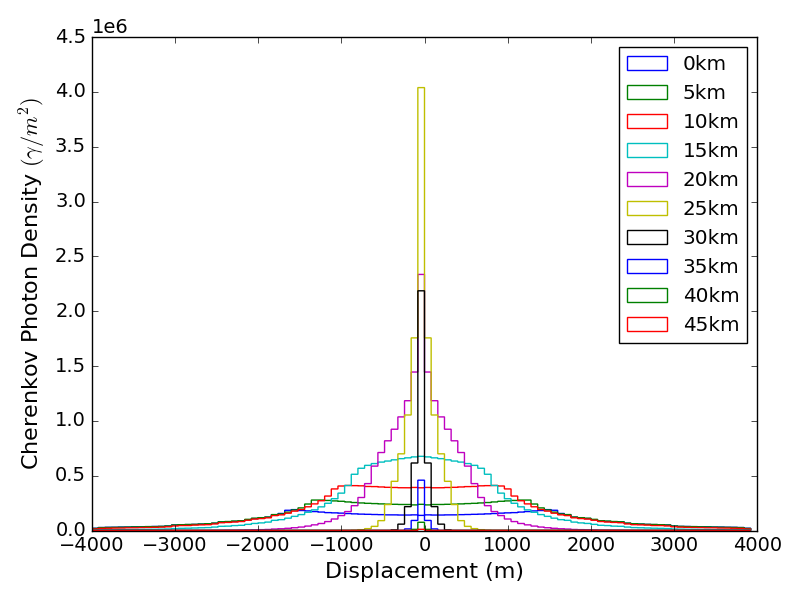} & 
  \includegraphics[width=.5 \textwidth]{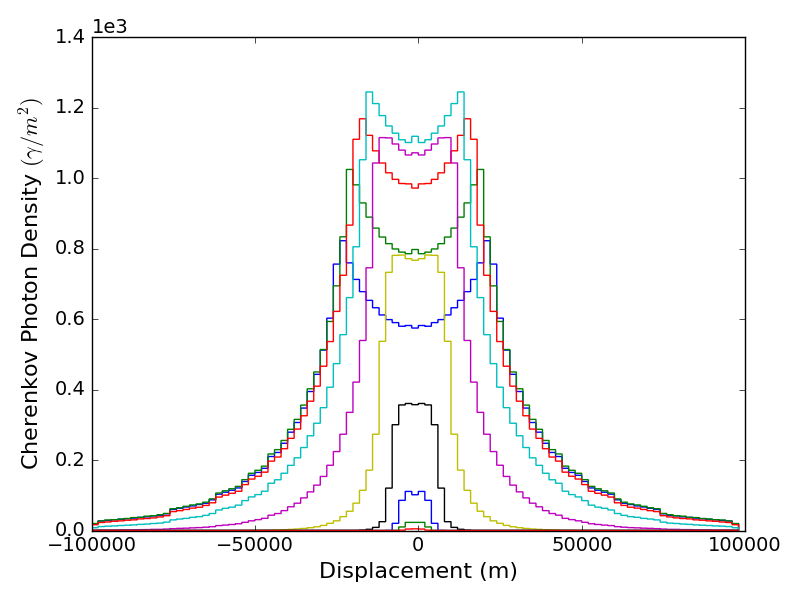}

\end{tabular}
 \caption{Cherenkov spatial distribution for a 100~PeV proton shower with $27^{\circ}$ Earth emergence angle (ANITA event 1) initiated at different decay distances for EUSO-SPB2 (left) and POEMMA (right)}
\label{Cherenkov_Spatial_Dists}
\end{figure}

For shower developments which occur lower in the atmosphere, the electron scale angle and Cherenkov angle are fairly similar near shower maximum, which would result in a significant fraction of the emitted Cherenkov photons being focused nearly uniformly inside the average Cherenkov angle. For showers which begin at ground, this results in a cone with near uniform photon density of radius 1.76~km for EUSO-SPB2 and 25~km for POEMMA. Outside the Cherenkov ring, intensity will decrease exponentially. When showers begin to develop deeper in the atmosphere, however, the Cherenkov angle shrinks due to decreased index of refraction, and the electron angular spread begins to dominate the overall behaviour. From this, we expect to see the resulting Cherenkov photon distribution depart from the double peaked structure and begin to follow the overall exponential behavior of the electrons. Additionally, the distance from emission point to detector plane decreases, resulting in an increase in intensity. With increasing tau decay distance, we expect to see a narrower, and brighter signal, becoming more exponential in behavior. When the shower encounters altitudes where the energy threshold for Cherenkov generation becomes too large, the intensity will begin to decrease. The spatial distribution for the ANITA like events with decay lengths between 0 and 50~km is plotted in figure \ref{Cherenkov_Spatial_Dists} for EUSO-SPB2 and POEMMA.

The spatial distribution for EUSO-SPB2 is significantly narrower than that of POEMMA, which is obvious, given their respective altitudes of 33~km and 525~km. This also explains the factor of roughly 1000 difference in their photon intensities. We observe the move to exponential behavior with increasing decay distance in both distributions, but we see it more clearly in the distribution of EUSO-SPB2. This can be explained by noting that the photons which develop at the greatest atmospheric depths geometrically spread much less than those at the beginning of the shower, given the shorter distance they must travel to the detector surface, giving them a greater effective weight in the distribution. The decay distance which produces the maximum Cherenkov photon density is different for the two instruments (25~km for EUSO-SPB2 versues 15~km for POEMMA). This is due to the fact that for some deeper shower developments, EUSO-SPB falls inside the shower development. This also explains why the behavior of the 25~km distribution for EUSO-SPB2 is very centralized around the shower axis.

\subsection{Photon Density}
\begin{figure}[t!]
\begin{tabular}{cc}
  \includegraphics[width=.5 \textwidth]{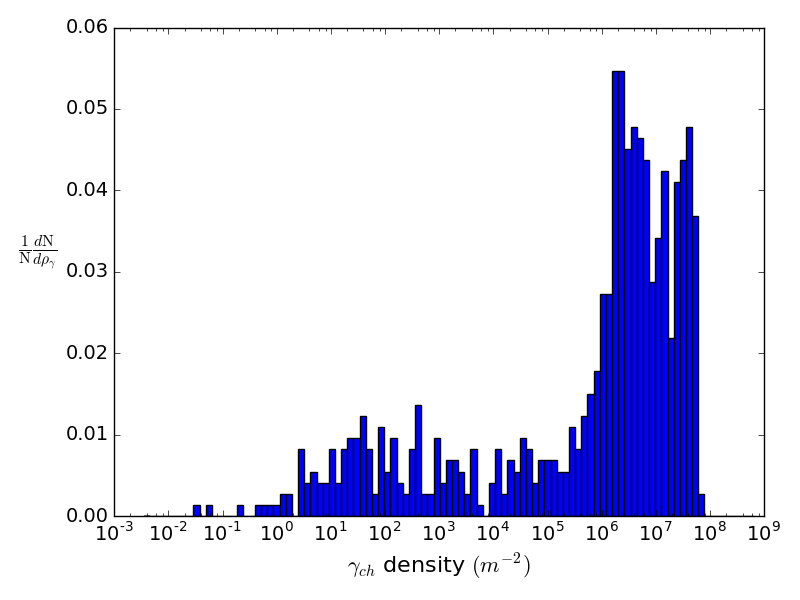} & 
  \includegraphics[width=.5 \textwidth]{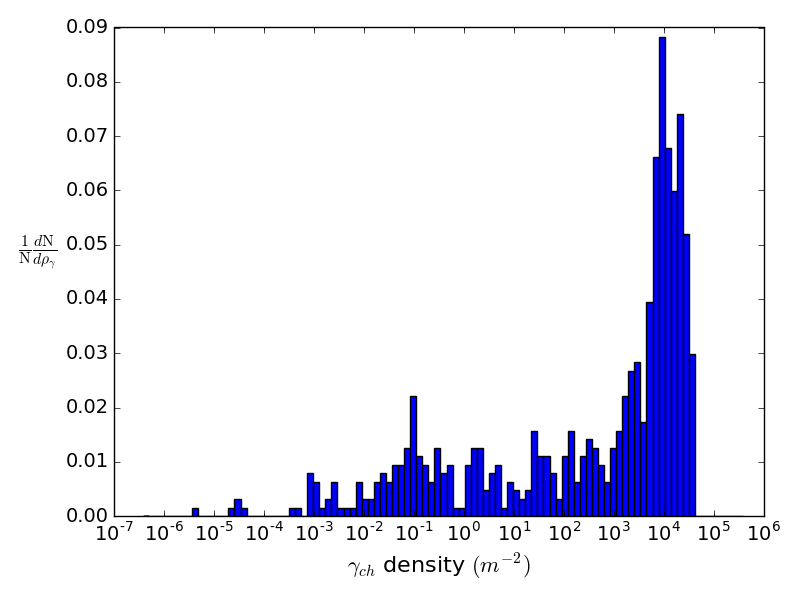}

\end{tabular}
 \caption{Distribution of central Cherenkov photon densities for 1000 1~EeV tau lepton showers with $27^{\circ}$ Earth emergence angle (ANITA event 1) for EUSO-SPB2 (left) and POEMMA (right)}
\label{Cherenkov_Intensities}
\end{figure}

We simulate 1000 1~EeV tau leptons, and distribute their starting locations in the atmosphere according to an exponential distribution with mean 50~km. Additionally, for these geometries, it is exceedingly unlikely that the muon decay branch of the tau lepton is observable through showering products, and therefore is neglected. We calculate the Cherenkov photon density only in the center of the shower axis (the central bin in figure \ref{Cherenkov_Spatial_Dists}). To reduce computing time, at each step, we assume $\frac{1}{5}$ of the total generated Cherenkov photons are distributed uniformly within the local Cherenkov angle. This approximation agrees fairly well with our spatial distributions described in the previous section, but we are working to develop a better parameterization \cite{me}. There exists a fraction of events which decay before the instrument, but too close to appreciably develop a visible Cherenkov profile. This fractions is roughly $20\%$ for both instruments. The central Cherenkov photon density for the ANITA like events is shown in figure \ref{Cherenkov_Intensities}.

A complete description of the airglow night background is beyond the scope of this proceedings paper, but has been taken into account by different authors \cite{neutrinos}. In order to reduce the rate of false positives from background, the photon density threshold of detection is set to $200 \gamma \mathrm{m}^{-2}$ and $20 \gamma \mathrm{m}^{-2}$, for EUSO-SPB2 and POEMMA, respectively \cite{EUSOSPB2} \cite{POEMMA}. The probability for an ANITA event to be above this photon threshold is $91\%$ and $73\%$.

\section{Discussion}
If the ANITA events are accurately described by a tau lepton initiated UAS, it is likely that an optical Cherenkov instrument such as EUSO-SPB2 or POEMMA would be able to observe them under the right circumstances. We have shown that the Cherenkov signals from such events would  be in the wavelength range of discovery, have signal strength well above background levels, and have a fairly wide spatial extent ($\mathcal{O}$(1~km) for EUSO-SPB2 and $\mathcal{O}$(10~km) for POEMMA). We have also shown that the decay distance of a tau lepton with ANITA-like energies is within the geometric range of discovery. Nevertheless, for both instruments to be sensitive to the pertinent Earth emergence angles of the ANITA events, slight design changes would be required in order to extend their angular range. A discussion is likely warranted into the weighting of costs, risks, and benefits for observing this angular region.

%We have shown here the reasons that the ANITA events cannot be described by conventional tau neutrino propagation and air showering using standard model approaches. However, there exist good explanations for these signals which circumvent explanations requiring exotic physics. Unfortunately, even if the explanations are accurate, radio signals alone cannot rule out the possibility of downward going showers appearing as upward going showers through various physical effects. Signatures for upward and downward going showers in the optical Cherenkov channel would likely appear appreciably different enough to make a clearer distinction. If the ANITA events are a result of radio emission from downward going showers exhibiting interesting interactions with ice, we would expect to see very dim signals because of increased geometric spread and atmospheric attenuation. But notably, if the ANITA events are evidence of beyond the standard model physics, an optical Cherenkov instrument such as EUSO-SPB2 or POEMMA would be sensitive enough to provide a very important and interesting confirmation.

We have shown here the reasons why the ANITA events cannot be described by conventional tau neutrino propagation and air showering using standard model approaches. There exist plausible explanations for these signals which circumvent explanations requiring exotic physics. Unfortunately, even if the explanations are accurate, radio signals alone cannot rule out the possibility of downward going showers appearing as upward going showers through various physical effects. Signatures for upward and downward going showers in the optical Cherenkov channel would likely appear appreciably different enough to make a clearer distinction. Notably, if the ANITA events are evidence of beyond the standard model physics, an optical Cherenkov instrument such as EUSO-SPB2 or POEMMA would be sensitive enough to provide a very important and interesting comparison

\end{document}